\documentstyle[aps,preprint,epsfig,eqsecnum]{revtex}
\begin{document}
\draft
\preprint{SNUTP/97-112, gr-qc/9803003}

\title{{\Large\bf Gravitating $\sigma$ Model Solitons}}
\author{Yoonbai Kim\thanks{Electronic mail address
: yoonbai$@$cosmos.skku.ac.kr}} 
\address{Department of Physics, Sung Kyun Kwan University, 
Suwon 440-746, Korea}
\author{Sei-Hoon Moon\thanks{Electronic mail address : jeollo$@$zoo.snu.ac.kr}}
\address{Department of Physics and Center for Theoretical Physics,\\ 
Seoul National University, 
Seoul 151-742, Korea}
\maketitle

\begin{abstract}
We study axially symmetric static solitons of O(3) nonlinear $\sigma$ model 
coupled to (2+1)-dimensional anti-de Sitter gravity. The obtained solutions 
are not self-dual under static metric. The usual regular topological lump
solution cannot form a black hole even though the scale of symmetry breaking
is increased. There exist nontopological solitons of half integral winding 
in a given model, and the corresponding spacetimes involve charged 
Ba$\tilde{\rm n}$ados-Teitelboim-Zanelli black holes without non-Abelian 
scalar hair.
\end{abstract}

\vspace{5mm}


\newpage

\section{Introduction}

Three-dimensional (3D) Einstein gravity is 
characterized by the absence of propagating gravitational degree
\cite{DJH}. Though it is different from the nature of (3+1)-dimensional 
gravity, 3D gravity without the graviton has attracted attention in
cosmology in connection with cosmic strings~\cite{VS} 
and in gauge theory formulation~\cite{AT}.
In both contexts, (2+1)-dimensional [(2+1)D] anti-de Sitter gravity may be 
intriguing because it was the first example reformulated as
a Chern-Simons gauge theory of the Poincar\'{e} group~\cite{AT} and 
its vacuum solutions support black holes~\cite{BTZ}.

(2+1)D gravity with a nonzero cosmological constant was first
studied
in Ref.~\cite{DJ}. When a static point particle with mass and without spin
is coupled to gravity, general anti-de Sitter solution was obtained
\begin{eqnarray}
\label{hype}
ds^2&=&\displaystyle{\sqrt{\varepsilon}
\frac{(\frac{R}{R_0})^{\sqrt{\varepsilon}~c}
+(\frac{R_0}{R})^{\sqrt{\varepsilon}~c}}{
(\frac{R}{R_0})^{\sqrt{\varepsilon}~c}
-(\frac{R_0}{R})^{\sqrt{\varepsilon}~c}}dt^2} 
\displaystyle{ - \frac{4\varepsilon c^2(dR^2+R^2d\Theta^2)}{
|\Lambda| R^2\Big[ (\frac{R}{R_0})^{\sqrt{\varepsilon}~c}
-(\frac{R_0}{R})^{\sqrt{\varepsilon}~c}
\Big]^2}},
\end{eqnarray}
where $c=1-4Gm$ and $\varepsilon$ is $\pm 1$ for the negative cosmological 
constant $\Lambda$.
When $\varepsilon=+1$, the metric (\ref{hype}) describes a hyperboloid 
with deficit angle.
Note that the effect of the point particle at the origin
appears only in the deficit angle
in Eq.~(\ref{hype}), and thereby these solutions go to vacuum solutions in
the
massless limit $(m\rightarrow 0)$. 
Later the Ban\~{a}dos-Teitelboim-Zanelli (BTZ) black hole solutions were
 reported in Ref.~\cite{BTZ}, and 
the simplest one is the Schwarzschild-type black hole
\begin{eqnarray}
\label{btzm}
ds^2=(|\Lambda|r^2-8GM)dt^2 -\frac{dr^2}{|\Lambda|r^2-8GM}-r^2d\theta^2.
\end{eqnarray}
Here an integration constant $M$ of Einstein equation is arbitrary, however
solutions of positive $M$ correspond to the BTZ black holes. 
Since both solutions in Eqs.~(\ref{hype})
and (\ref{btzm}) are vacuum solutions in the limit of zero point particle 
mass, one may easily find a coordinate transformation to connect 
the $m=0$ solutions in Eq.~(\ref{hype}) with the solutions in Eq.~(\ref{btzm}). 
As expected,
$\varepsilon=+1$ case in Eq.~(\ref{hype}) corresponds to the negative
$M$ solution in Eq.~(\ref{btzm}), and the corresponding space is 
a regular hyperboloid.
 $\varepsilon=-1$ case results in the 
exterior region of the Schwarzschild-type BTZ black hole~\cite{KKK}.

This BTZ black hole has so far attracted much interest in various  classical
black hole solutions \cite{Cle}, in thermodynamic and statistical properties
\cite{Ther,Car}, and in string related topics \cite{Stri}.
In 3+1 dimensions, gravitating solitons and sphalerons have received
considerable impetus by the discovery of a class of non-Abelian black hole
solutions \cite{BM,LNW,Wei}. It might be an intriguing direction to ask the 
same question that whether or not gravitating solitons in (2+1)D anti-de Sitter
spacetime can form solitonic BTZ black holes. 
In case of global U(1) vortices, a regular
configuration could make a black hole structure with two horizons similar to
the charged BTZ black hole \cite{KKK}. Since the energy of a static global
U(1) vortex diverges logarithmically in flat spacetime, we here want to address
the same question to a model containing finite energy soliton excitations.
In this context 
O(3) nonlinear $\sigma$ model may be an appropriate choice 
since the field content
of the model is simple, and exact static self-dual multi-soliton solutions
of finite energy have been obtained in both flat
and curved spacetime with zero cosmological constant \cite{CG,GOR1,HP}.

In this paper, we consider both negative cosmological constant and 
matter distribution provided by regular static solitons of O(3) 
nonlinear $\sigma$ model. The metric of our consideration is static 
and axially symmetric. 
The inclusion of a negative cosmological constant makes us expect to induce
drastic change to solitonic physics in 2+1 dimensions. A role of it is
effectively
equivalent to the introduction of angular momentum under a stationary metric, 
and then the corresponding
spacetime provides a rotating frame to the test particle.
Therefore, static $\sigma$ solitons in anti-de Sitter spacetime cannot remain
to be self-dual under the static metric. 
Even if we obtain the self-dual $\sigma$ solitons under the stationary metric,
we encounter unphysical situation,
e.g., closed timelike curves \cite{KK}.
Attractive gravitational force 
sounds natural in 3+1 dimensions for localized ordinary matter distributions, 
so that it makes the matter collapse into the black
hole or coagulates a new localized object which does not exist in flat
spacetime \cite{BM}. Since (2+1)D gravity itself does not contain propagating 
gravitational field, negative vacuum energy can induce a similar effect in
curved spacetime. In O(3) nonlinear $\sigma$ model, we present a new
nontopological soliton solution of half integral winding in addition to the
well-known topological lump solution of integral winding. We also show that
any regular topological lump whose energy is localized near its core cannot 
form spacetime of a BTZ black hole. However, the nontopological solutions
have a logarithmically divergent energy tail, so that their spacetimes can
include
charged BTZ black hole. In these aspects the obtained nontopological solitons 
resemble global U(1) vortices, but the non-Abelian scalar hair of 
$\sigma$ solitons
do not penetrate the horizon while the scalar hair of the 
global U(1) vortices can be
observed outside the BTZ black hole.
 
This paper
is organized as follows. In section II, we introduce the model and obtain all
possible static regular solitons with axial symmetry by solving  
second order Euler-Lagrange equations. In section III, 
the spacetime structure including BTZ black holes is analyzed for the
obtained gravitating solitons. Geodesic motions are
computed in Sec. IV. We conclude in Sec. V
with a discussion.

\section{Model and Soliton Solutions}

\indent
Nonlinear $\sigma$ model with O(3) symmetry is described by the Lagrange density
\begin{eqnarray}\label{act}
{\cal L}= -\frac{1}{16\pi G}(R+2\Lambda)
+\frac{1}{2}g^{\mu\nu}\partial_\mu {\phi}^a \partial_\nu {\phi}^a 
-\frac{\lambda(x)}{2}v^{2}({\phi}^a{\phi}^a - v^{2} ),
\end{eqnarray}
where a Lagrange multiplier $\lambda(x)$ is rescaled to a dimensionless 
quantity,
and the variation of it produces a constraint for the scalar field:
$\phi^{a}\phi^{a}=v^{2}\;(a=1,2,3)$.
Throughout this paper, the dimension counting of fields is adjusted
to that in (3+1)-dimensional spacetime since we presume to apply the obtained 
results to the 
straight, infinite strings. Then the model involves three mass scales, 
namely the Planck scale
$1/\sqrt{G}$, the scale of negative cosmological vacuum energy
$\sqrt{|\Lambda|}$, and the symmetry breaking scale $v$.
Solitonic objects of our interest have axial symmetry,  
i.e., the corresponding string spacetime is invariant under the rotation to,
and the translation along a symmetry axis. 
The mass in this paper stands for mass per unit length along the symmetry
axis.
In this case the static metric of
this spacetime can be parametrized as 
\begin{eqnarray}
\label{cyl}
ds^2=e^{2N(r)}B(r)dt^2-\frac{dr^2}{B(r)}-r^2d\theta^2-dz^2. 
\end{eqnarray}
For this kind of the metric all physical settings are effectively reduced the 
hypersurface orthogonal to the symmetry axis, and the string-like object
can be viewed as a point-like source in 2+1 dimensions.
Suppose that a given matter distribution is specialized to the case of axially
symmetric time-independent fields and the equations of motions are solved.
The resulting metric has two integration constants that are identified as the
mass and angular momentum \cite{BTZ}. Since we take a static metric
(\ref{cyl}) here, it is equivalent to set the angular momentum zero. 
When we fix the boundary condition at the origin for the fields and the metric, 
we will choose a value of the mass parameter $B(0)$ later.
We take a stereographic projection for $\phi^{a}$ so that
the ansatz for the solitons with axial symmetry is 
\begin{eqnarray}\label{stereo}
{\phi}^a = v(\sin F(r)\cos n\theta, \sin F(r)\sin n\theta, \cos F(r) ). 
\end{eqnarray}
Euler-Lagrange
equations derived from the action and the static metric are 
\begin{eqnarray}\label{eleq}
\frac{d^{2}F}{dr^{2}} + \Bigl( \frac{dN}{dr}
          + \frac{1}{B} \frac{dB}{dr}
         + \frac{1}{r} \Bigr)\frac{dF}{dr}
         = \frac{n^2}{ B r^2}\sin F \cos F ,
\end{eqnarray}
\begin{eqnarray}\label{Neq}
\frac{1}{r}\frac{dN}{dr}= 8 \pi G v^{2}\Bigl( \frac{dF}{dr} \Bigr)^2 ,
\end{eqnarray}
\begin{eqnarray}\label{Beq}
\frac{1}{r}\frac{dB}{dr}=2|\Lambda|
       - 8\pi G v^{2}\biggl\{ B\Bigl( \frac{dF}{dr} \Bigl)^{2} + 
       \frac{n^2}{r^2} \sin^2 F \biggr\}.
\end{eqnarray}

A physical condition for spacetime manifold is the reproduction
of Minkowski spacetime
in the limit of no matter ($T^{\mu}_{\;\nu}=0$) and zero cosmological 
constant ($\Lambda=0$), and then  
an appropriate set of boundary conditions is
\begin{eqnarray}\label{bczero}
B(0)=1 ~~~{\rm and}~~~N(\infty)=0. 
\end{eqnarray}
When  $n\neq 0$, well-definedness of the scalar field $\phi^a$ in 
Eq.~(\ref{stereo}) forces the boundary condition at the origin such as
\begin{eqnarray}\label{fbc}
F(0)=0~~~~~~~(\mbox{or}~~\sin F(0)=0).
\end{eqnarray}
Introducing a new variable $\tilde{r}=\ln r~~(-\infty < \tilde{r}<\infty) $, 
we rewrite Eq.~(\ref{eleq}) such as
\begin{eqnarray}\label{eleqr}
\frac{d^2F}{d\tilde{r}^2}+\Bigl( \frac{dN}{d\tilde{r}}+\frac{1}{B}
\frac{dB}{d\tilde{r}} \Bigr)\frac{dF}{d\tilde{r}}= \frac{n^2}{B}\sin F\cos F.
\end{eqnarray}
After eliminating derivative terms of the metric functions by use of 
Eqs.~(\ref{Neq}) and (\ref{Beq}), we obtain 
\begin{eqnarray}\label{Newt}
B\frac{d^2F}{d\tilde{r}^2}=n^2\sin F \cos F - (2|\Lambda|e^{2\tilde{r}}
-8\pi Gv^2n^2\sin^2 F)\frac{dF}{d\tilde{r}}.
\end{eqnarray}
{}From the vanishment of the right-hand side of Eq.~(\ref{eleqr}) at spatial
infinity, we read
possible boundary values of the scalar amplitude:
\begin{eqnarray}\label{fbdinf}
F(\infty)=\left\{
          \begin{array}{ll}
          \pi~~ &\mbox{from the sine term,}  \\
          \pi/2 ~~&\mbox{from the cosine term,} \\
          \alpha\;(0<\alpha\leq\pi)~~&\mbox{from}\;1/B(\infty)\;\mbox{term}.
          \end{array}\right.
\end{eqnarray}
The boundary condition in the last line of Eq.~(\ref{fbdinf}) comes from the
divergence of $B(r)$ at spatial infinity.
Precisely, $B(r)\approx |\Lambda|r^2$ for a sufficiently large $r$.

Before analyzing $n\neq 0$ solutions of Eq.~(\ref{eleq}), we will show
that there does not exist $n=0$ regular nontrivial solution of this 
equation even in anti-de Sitter space.
If we substitute Eqs.~(\ref{Neq}) and (\ref{Beq}) into Eq.~(\ref{eleq})
when  $n=0$, we obtain 
\begin{eqnarray}\label{neq}
\frac{d^2F}{dr^2} + \Bigl(\frac{2\Lambda r}{B}+\frac{1}{r}\Bigr)
\frac{dF}{dr}=0.
\end{eqnarray}
Since $B(0)=1$, $F$ given by a solution of this equation contains 
a logarithmic divergence
at the origin, i.e.,
$F(r)\propto \int dr^2 e^{-|\Lambda|r^2}/r^2$
for a sufficiently small $r$.
Now that we have shown nonexistence of the $n=0$ solution, 
let us look for the 
$n\neq 0$ soliton solutions of the equations 
(\ref{eleq}), (\ref{Neq}), and (\ref{Beq}) satisfying the boundary
conditions in Eqs.~(\ref{bczero}), (\ref{fbc}), and (\ref{fbdinf}).

\subsection*{\bf{II.1\,\,\, Topological Soliton}}

Solutions satisfying the boundary condition that $F(0)=0$ and $F(\infty)=
\pi$ are topological solitons when the base spatial manifold formed by them 
is topologically equivalent to two dimensional Euclidean space. These static
solitons are characterized by topological charge,
\begin{eqnarray}\label{topch}
Q&=&\frac{1}{8\pi}\int d^{2}x\epsilon^{0ij}\epsilon^{abc}\phi^{a}\partial_{i}
\phi^{b}\partial_{j}\phi^{c}, \\
&=&\frac{n}{2}(\cos F(0)-\cos F(\infty)), \\
&=&n,
\end{eqnarray}
and this quantized charge $n$ represents a winding number of second
homotopy group, that is, $\Pi_{2}(S^{2})=Z$. 
From now on we will call topological solitons
of this model `topological lumps'.

The topological lumps are known to be unique static soliton species of
O(3) nonlinear $\sigma$ model in flat spacetime, and they have been studied in
curved spacetime as a candidate of global cosmic strings \cite{CG,GOR1,HP}.
Since the exact soliton solutions were obtained by solving first order
self-dual equation,
their existence has been automatic as far as the cosmological constant has not 
been taken into account. As we shall discuss it later, static solitons 
under the static metric are not self-dual in anti-de
Sitter spacetime and then we have to consider the second order Euler-Lagrange
equation (\ref{eleq}) directly.

Since we cannot exactly solve the equations (\ref{eleq}), (\ref{Neq}), 
and (\ref{Beq}),
let us attempt series expansion of the fields near the origin
\begin{eqnarray}
F(r)&\approx&F_0 r^n -\big(|\Lambda|-8\pi Gv^{2}F_{0}^{2}+
F^{3}_{0}/2\big) r^{3}, \label{fzero}\\
N(r)&\approx&N_0+4\pi Gv^2F_0^2 n r^{2n}, \label{nzero}\\
B(r)&\approx& 1+\Bigl[\frac{|\Lambda|}{v^{2}}-4\pi G
                  (1+n^2)F_0^2 \delta_{1,n}\Bigr](vr)^2,\label{bzero}
\end{eqnarray}
where $F_{0}$ and $N_{0}$ are constants determined by proper behavior of
the fields at asymptotic region.
For large $r$ the leading term approximation gives
\begin{eqnarray}
F(r)&\approx&\pi - \frac{F_{\infty}}{r^2},
\label{finf}\\
N(r)&\approx&-\frac{8\pi G v^2F_{\infty}^2}{ r^4},\label{ninf}\\
B(r)&\approx& |\Lambda| r^2 +B_{\infty}+ \frac{16\pi Gv^2 |\Lambda|F_{\infty}^2
}{r^2},\label{binf}
\end{eqnarray}
where $F_{\infty}$ and $B_{\infty}$ are also determined by the proper 
functional behavior at the origin.

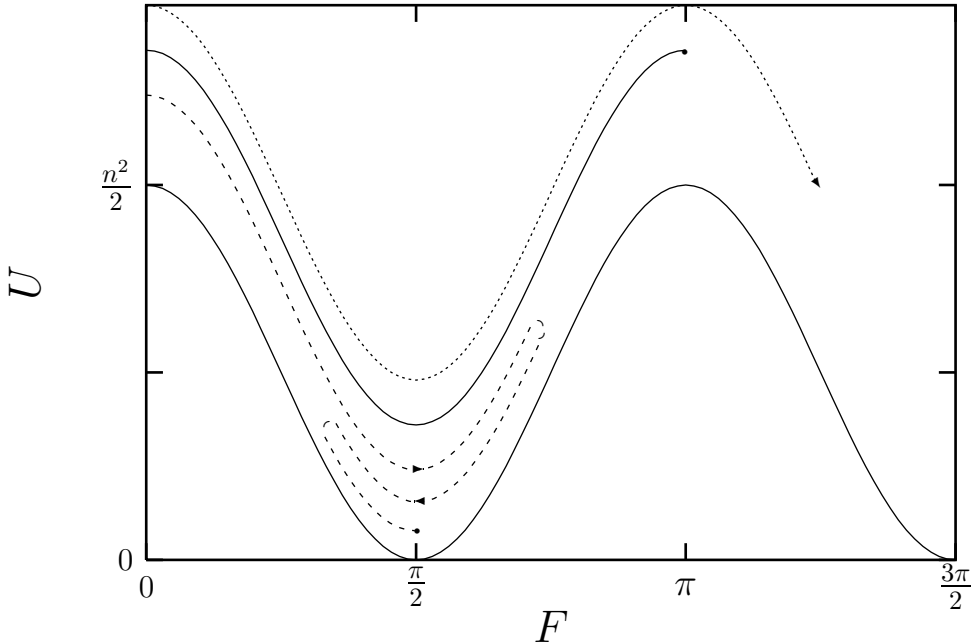
\begin{figure}

\setlength{\unitlength}{0.1bp}
\begin{picture}(3600,2592)(0,0)
\put(2025,150){\makebox(0,0){{\Large $F$}}}
\put(100,1446){%
\makebox(0,0)[b]{\shortstack{{\Large $U$}}}%
}
\put(3550,300){\makebox(0,0){{\large $\frac{3\pi}{2}$}}}
\put(2533,300){\makebox(0,0){{\large $\pi$}}}
\put(1517,300){\makebox(0,0){{\large $\frac{\pi}{2}$}}}
\put(500,300){\makebox(0,0){0}}
\put(450,1814){\makebox(0,0)[r]{{\large $\frac{n^{2}}{2}$}}}
\put(450,400){\makebox(0,0)[r]{0}}
\put(2539,2317){\makebox(0,0){\circle*{20}}}
\put(1530,507){\makebox(0,0){\circle*{20}}}
\put(1510,620){\makebox(0,0){\vector(-1,0){1}}}
\put(1545,792){\makebox(0,0){\vector(1,0){1}}}
\put(3040,1852){\makebox(0,0){\vector(2,-3){1}}}
\put(1170,920){\makebox(0,0){\oval(2,2)[tl]}}
\put(1993,1300){\makebox(0,0){\oval(3,3)[tr]}}
\put(1999,1255){\makebox(0,0){\oval(2,2)[br]}}
\end{picture}

\vspace{3mm}

\caption{Shape of the effective potential $U$ and possible motions of 
a hypothetical particle: (a) overshoot solution (the dotted line), (b)
critical solution with $F(\infty)=\pi$ (the solid line), (c) undershoot
solution with $F(\infty)=\pi/2$ (the dashed line).}
\label{fig1}
\end{figure}

If we identify $F$ as a coordinate
and $\tilde{r}$ as time in Eq.~(\ref{Newt}), then we can interpret this
equation as a Newtonian
equation for one-dimensional motion of a hypothetical particle with
variable mass $B(r)$. The exerted forces are friction or a kind of
velocity-dependent force proportional to $dF/d\tilde{r}$,
and the conservative force from the potential
$U=\frac{n^2}{2}\cos 2F$ (See Fig.~\ref{fig1}).

If we naively read possible motions of a hypothetical particle from
the potential $U(F)$, then the motions satisfying
$F(r=0)=0$ are classified
into three sets by its initial velocity which can actually be replaced by
the value of $F_0$ in 
Eq.~(\ref{fzero}). When $F_0$ is larger than a critical value, the 
particle reaches $\pi$ at a finite time $\tilde{r}$ and it corresponds to an 
overshoot shown by the dotted line in Fig.~\ref{fig1}.  
When $F_0$ is smaller than the critical value, the particle cannot reach
$\pi$ because of the power loss due to the velocity-dependent terms
in Eq.~(\ref{Newt}) and this motion should have a turning point between $\pi/2$ 
and $\pi$.
The existence of the overshoot solution given by the dotted line in
Fig.~\ref{fig1} and the undershoot solution
given by the dashed line in Fig.~\ref{fig1}
guarantees, by continuity argument, the existence of the 
topological lump solution connecting $F(r=0)=0$ and $F(r=\infty)=\pi$
smoothly (See the solid line in Fig.~\ref{fig1}). 

For the metric functions, $N(r)$ is monotonically increasing since the 
right-hand side of Eq.~(\ref{Neq}) is always nonnegative, however 
$N(r)$ is slowly varying function in the asymptotic region as was shown
in Eq.~(\ref{ninf}). It means that the exponential of $N(r)$ 
does not affect 
much 
to the structure
of spacetime. On the other hand, functional behavior of $B(r)$ 
changes drastically according to both the magnitude of the cosmological 
constant and the matter
distribution.
Therefore, its spacetime structure, e.g., a black hole, is determined by 
reading the shape of 
$B(r)$. 
We will discuss possible spacetimes generated by various 
$\sigma$ solitons in the next section.

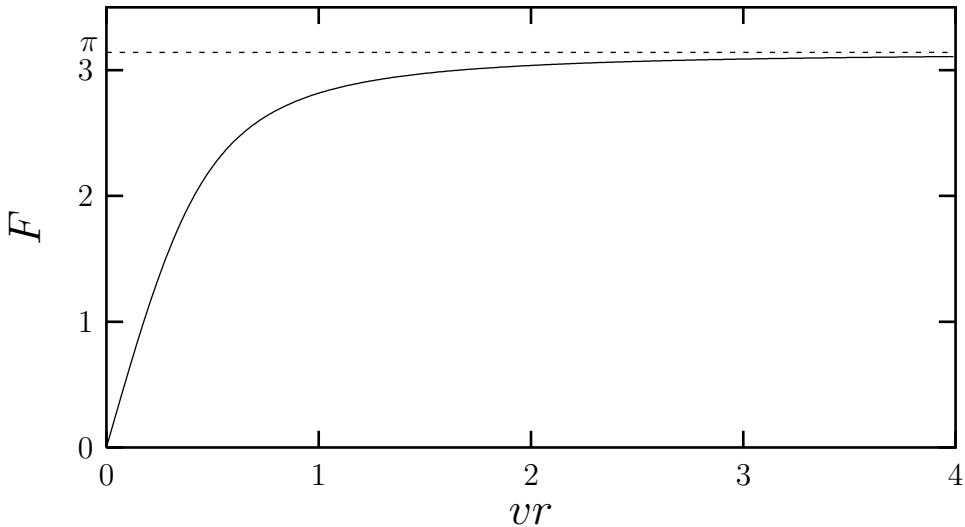
\begin{figure}


\setlength{\unitlength}{0.1bp}
\begin{picture}(3600,2160)(0,0)
\put(1950,150){\makebox(0,0){{\Large $vr$}}}
\put(100,1230){%
\makebox(0,0)[b]{\shortstack{{\Large $F$}}}%
}
\put(3550,300){\makebox(0,0){4}}
\put(2750,300){\makebox(0,0){3}}
\put(1950,300){\makebox(0,0){2}}
\put(1150,300){\makebox(0,0){1}}
\put(350,300){\makebox(0,0){0}}
\put(320,1923){\makebox(0,0)[r]{$\pi$}}
\put(300,1823){\makebox(0,0)[r]{3}}
\put(300,1349){\makebox(0,0)[r]{2}}
\put(300,874){\makebox(0,0)[r]{1}}
\put(300,400){\makebox(0,0)[r]{0}}
\end{picture}

\vspace{3mm}

\caption{A configuration of topological lump solution when $8\pi Gv^{2}=0.2$,
$|\Lambda|/v^{2}=4.0\times 10^{-6}$, and $F_{0}=5.896$. 
The boundary value of the
topological lump solution has $\pi$ with $10^{-6}$ precision.}
\label{fig2}
\end{figure}

In the above discussion, we neglected the effect of the variable mass
$B(r)$ in Eq.~(\ref{Newt}). It may be valid when the absolute value of the 
cosmological constant is small. 
On the other hand, if $|\Lambda|/v^{2}$ is large enough, 
the terms proportional to the cosmological constant dominate even for some
finite $\tilde{r}$ region. In the Newtonian equation (\ref{Newt}), such terms
are interpreted as the variable mass term 
$B(\tilde{r})\sim|\Lambda|e^{2\tilde{r}}$ and
the time-dependent coefficient of the friction $2|\Lambda|e^{2\tilde{r}}$
in the right-hand side of Eq.~(\ref{Newt}), respectively.
In this case, the mass of 
the hypothetical particle can rapidly increase for small $r$ and it can
forbid the existence of overshoot solutions even for huge $F_0$ values.
It is indeed the case which was confirmed by numerical computation.
In synthesis, there exists regular topological lump solution satisfying
the boundary conditions, $F(0)=0$ and $F(r=\infty)=\pi$, only when
$|\Lambda|/v^2$ is less than a critical value. 
An example of the topological lump is shown in Fig.~\ref{fig2}.

\subsection*{\bf{II.2\,\,\,Nontopological Soliton}}

When we discussed solutions of Eq.~(\ref{fbdinf}) in the previous subsection,
we discussed possibility of another set of regular
solution satisfying
$F(\infty)=\alpha$ $(0<\alpha<\pi)$ as given in Eq.~(\ref{fbdinf}). 
Suppose that there exist
such solutions and we attempt power series expansion of them for large $r$:
\begin{eqnarray}\label{falpha}
F(r)\sim \alpha - \frac{F_{\alpha,\infty}}{r^q}. 
\end{eqnarray}
{}From Eqs.~(\ref{Neq}) and (\ref{Beq}), we have 
\begin{eqnarray}
& &N(r)\sim -4\pi Gv^2q\frac{F^2_{\alpha,\infty}}{r^{2q}},
\label{nalpha}\\
& &B(r)\sim|\Lambda|r^2 +1-8G{\cal M}_{\alpha}-
8\pi Gv^2n^2\sin^2\alpha \ln r/r_{\tilde{c}},\label{balpha}
\end{eqnarray}
where $F_{\alpha,\infty}$ and ${\cal M}_{\alpha}$ are constants 
which have to be chosen by the proper behavior of $F(r)$ and $B(r)$ near
the origin, and $r_{\tilde{c}}$ stands for core radius. 
Inserting the series solutions (\ref{falpha}), (\ref{nalpha}), and 
(\ref{balpha})
into the equation (\ref{eleq}) of the scalar field, we have a relation for 
the leading term
\begin{eqnarray}\label{expan}
-q(q-2)\frac{|\Lambda|F_{\alpha,\infty}}{r^{q}}=\frac{n^{2}}{r^{2}}
\sin\alpha\cos\alpha .
\end{eqnarray}
When $\alpha\neq\pi/2$ and $0<\alpha<\pi$, the functional behavior of
the radial coordinate forces $q=2$ but then the equality 
cannot hold because of the vanishment of the left-hand side of 
Eq.~(\ref{expan}). This implies impossibility of 
regular $F(\infty)=\alpha$ 
solution except $F(\infty)=\pi/2$ solution.
When the boundary value of $F$ is $\pi/2$, the charge defined in
Eq.~(\ref{topch}) is a multiple of half, i.e., $Q=n/2$. 
Therefore, every solution of $F(\infty)=\pi/2$ is classified as a static 
nontopological soliton of half integral winding.

In the previous subsection we mentioned existence of undershoot
solutions, and they should be nothing but the solutions of $F(\infty)=\pi/2$. 
Here let us emphasize again the impossibility of 
this half integral winding
solution in flat spacetime. Since $N(r)=0$ and $B(r)=1$ in flat spacetime,
Eq.~(\ref{eleqr}) depicts a one-dimensional motion of a hypothetical
particle with unit mass of which position is $F$ at time $\tilde{r}$.
The exerted force comes only from the conservative potential $U(F)$
shown in Fig.~\ref{fig1}, so virial theorem allows two regular solutions, 
i.e., the stopped motion $( F(\tilde{r})=0 )$ or the motion satisfying 
$F(\tilde{r}=-\infty)=0$ and $F(\tilde{r}=\infty)=\pi$. 
In curved spacetime with zero cosmological constant, the
velocity-dependent force is not a friction but it pushes the hypothetical
particle outward. Moreover the variable mass $B(r)$ of the particle 
decreases as time $\tilde{r}$ elapses. These two factors make turning of 
the hypothetical particle more difficult before $F=\pi$ and forbid 
undershoot solution.
Therefore, there does not exist any nontopological solitons of half integral
winding in curved spacetime when the cosmological
constant vanishes. In de Sitter spacetime, the positive cosmological 
constant term
makes the situation worse, so we easily expect no half integral winding 
solution similar
to the case of zero cosmological constant. In anti de Sitter spacetime,
the negative cosmological constant term provides a friction as shown 
in Eq.~(\ref{Newt}) and lets the variable mass $B(r)$ get heavy for large
$r$ as given in Eq.~(\ref{binf}). Among the solutions classified by the
value of $F_0$ in Eq.~(\ref{fzero}), a set of $F_0$'s less than the critical
value for the topological lump solution provides a set of 
undershoot solutions with turning point between $\pi/2$ and $\pi$.
Since the potential $U$ has minimum at $\pi/2$, it may oscillate around 
$\pi/2$ and finally converges to $\pi/2$ due to the friction. 

For better understanding of the asymptotic behavior of the scalar field $F(r)$,
let us consider linearized equation for $\delta F(r)$ defined by
$F(r)=\pi/2+\delta F(r)$. As an approximation of $B(r)$ we bring up two cases:
One describes the region of slowly varying $B$ ($B(r)\approx \bar{B}$),
and the other is the asymptotic region ($B(r)\approx |\Lambda|r^{2}$). 
The former leads to
\begin{eqnarray}\label{lineq1}
\bar{B}\frac{d^2\delta F}{dr^2}+3|\Lambda|r\frac{d\delta F}{dr}+
\frac{n^2}{r^{2}}\delta F=0,
\end{eqnarray}
and the latter goes to
\begin{eqnarray}\label{lineq2}
|\Lambda|r^2\frac{d^2\delta F}{dr^2}+3|\Lambda|r\frac{d\delta F}{dr}+
\frac{n^2}{r^{2}}\delta F=0.
\end{eqnarray}
A representative asymptotic solution of each equation is given in 
Fig.~\ref{fig3}
and every solution includes both oscillation and damping as expected. 
Note that
oscillations are rapid for small $r$ but the period of each oscillation
also increases rapidly as $r$ increases. Since this small $r$ region of rapid
oscillation is covered by the soliton core, we may expect possibility of
monotonic solution. It is indeed a case and we obtain a class of solutions
specified by the number of $\pi/2$ points at finite $r$. 
{}From now on we will call this
number as ``node''. From the value of $F_{0}$ in Fig.~\ref{fig4} one may
easily read proportionality between $F_{0}$ and the nodes. Obviously the
maximum value of $F$ also increases as $F_{0}$ becomes larger.

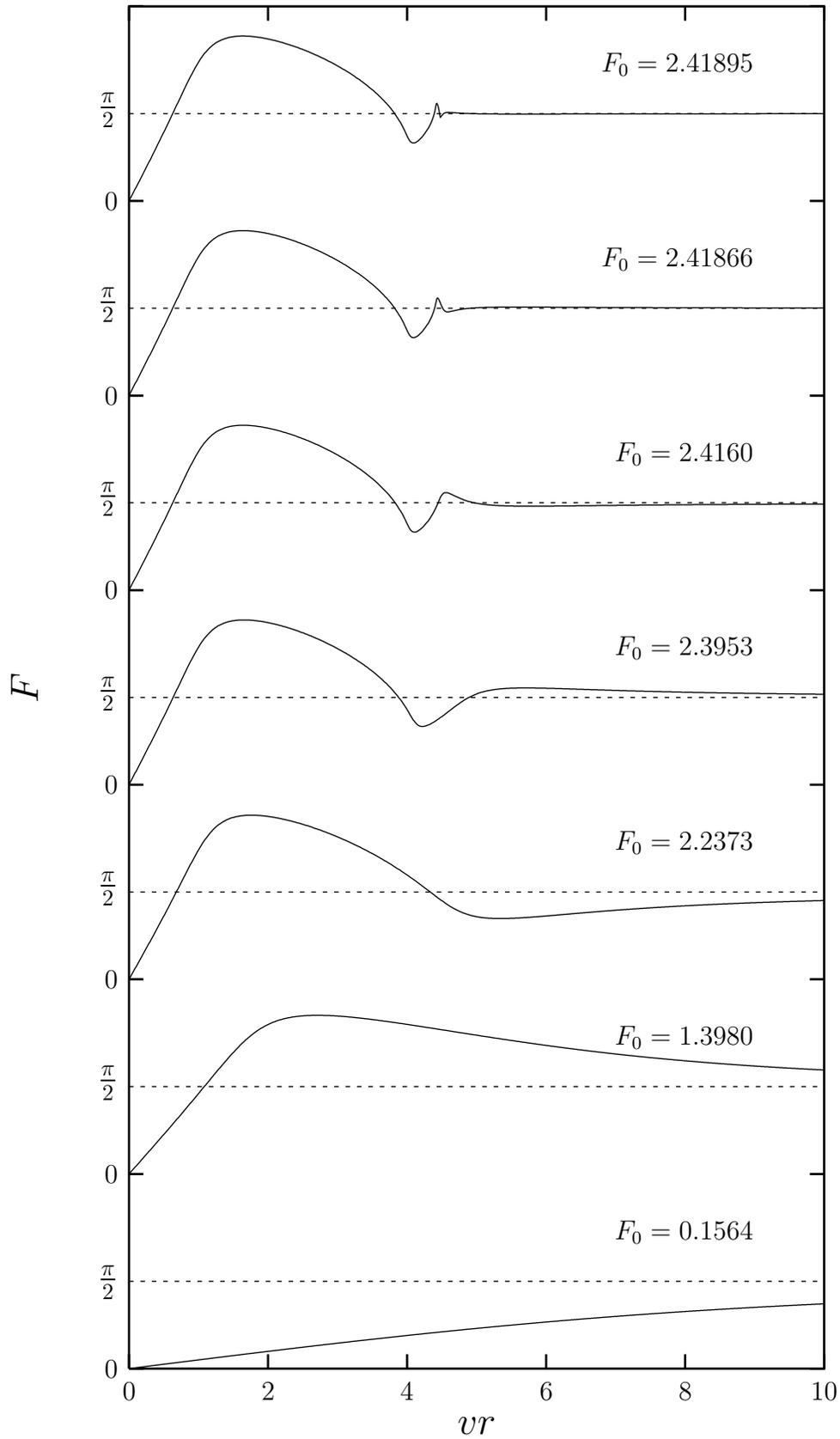
\begin{figure}

\setlength{\unitlength}{0.1bp}
\begin{picture}(3600,2160)(0,0)
\put(2025,150){\makebox(0,0){{\Large $vr$}}}
\put(100,1230){%
\makebox(0,0)[b]{\shortstack{{\Large $F$}}}%
}
\put(3550,300){\makebox(0,0){10}}
\put(2940,300){\makebox(0,0){8}}
\put(2330,300){\makebox(0,0){6}}
\put(1720,300){\makebox(0,0){4}}
\put(1110,300){\makebox(0,0){2}}
\put(500,300){\makebox(0,0){0}}
\put(450,2060){\makebox(0,0)[r]{1}}
\put(450,1645){\makebox(0,0)[r]{0.5}}
\put(450,1230){\makebox(0,0)[r]{0}}
\put(450,815){\makebox(0,0)[r]{-0.5}}
\put(450,400){\makebox(0,0)[r]{-1}}
\end{picture}

\vspace{3mm}

\caption{Two types of asymptotic solutions for $\delta F(r)\equiv
F(r)-\pi$ when $8\pi Gv^{2}=0.4$ and $|\Lambda|/v^{2}=0.01$.
Dashed line is a solution of Eq.~(\ref{lineq1}) when
$F_{0}=0.15$, and $F(r=0.01)=0.0001$.
Solid line is a solution of Eq.~(\ref{lineq2}) when
$F_{0}=10$, and $F(r=0.3)=-1$.}
\label{fig3}
\end{figure}
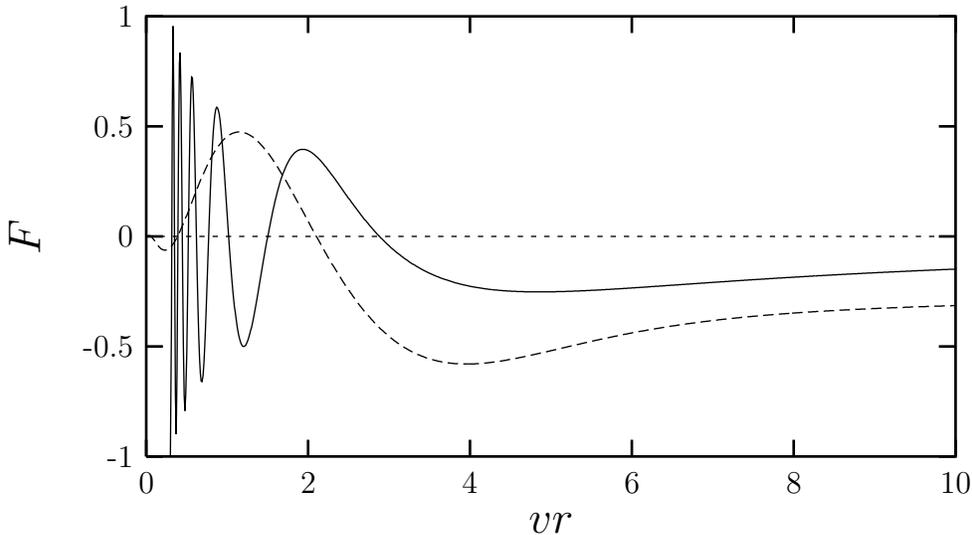

\begin{figure}

\setlength{\unitlength}{0.1bp}
\begin{picture}(3600,6480)(0,0)
\put(3237,6126){\makebox(0,0)[r]{$F_{0}=2.41895$}}
\put(3237,5271){\makebox(0,0)[r]{$F_{0}=2.41866$}}
\put(3237,4417){\makebox(0,0)[r]{$F_{0}=2.4160$}}
\put(3237,3563){\makebox(0,0)[r]{$F_{0}=2.3953$}}
\put(3237,2709){\makebox(0,0)[r]{$F_{0}=2.2373$}}
\put(3237,1854){\makebox(0,0)[r]{$F_{0}=1.3980$}}
\put(3237,1000){\makebox(0,0)[r]{$F_{0}=0.1564$}}

\put(2025,150){\makebox(0,0){{\Large $vr$}}}
\put(100,3390){%
\makebox(0,0)[b]{\shortstack{{\Large $F$}}}%
}
\put(3550,300){\makebox(0,0){10}}
\put(2940,300){\makebox(0,0){8}}
\put(2330,300){\makebox(0,0){6}}
\put(1720,300){\makebox(0,0){4}}
\put(1110,300){\makebox(0,0){2}}
\put(500,300){\makebox(0,0){0}}
\put(450,5526){\makebox(0,0)[r]{0}}
\put(450,4671){\makebox(0,0)[r]{0}}
\put(450,3817){\makebox(0,0)[r]{0}}
\put(450,2963){\makebox(0,0)[r]{0}}
\put(450,2109){\makebox(0,0)[r]{0}}
\put(450,1254){\makebox(0,0)[r]{0}}
\put(450,400){\makebox(0,0)[r]{0}}
\put(450,5926){\makebox(0,0)[r]{{\large $\frac{\pi}{2}$}}}
\put(450,5071){\makebox(0,0)[r]{{\large $\frac{\pi}{2}$}}}
\put(450,4217){\makebox(0,0)[r]{{\large $\frac{\pi}{2}$}}}
\put(450,3363){\makebox(0,0)[r]{{\large $\frac{\pi}{2}$}}}
\put(450,2509){\makebox(0,0)[r]{{\large $\frac{\pi}{2}$}}}
\put(450,1654){\makebox(0,0)[r]{{\large $\frac{\pi}{2}$}}}
\put(450,780){\makebox(0,0)[r]{{\large $\frac{\pi}{2}$}}}
\end{picture}

\vspace{3mm}

\caption{Various nontopological solitons specified by the number of nodes when 
$8\pi Gv^{2}=0.4$ and $|\Lambda|/v^{2}=0.01$.}
\label{fig4}
\end{figure}

Now some comments on $B(r)$ for large $r$ are in order. The expression
(\ref{balpha}) involves logarithmic term when $\alpha=\pi/2$, and it means
resemblance between the obtained nontopological solitons of half integral
 winding and
the vortices in a scalar model with global U(1) symmetry \cite{KKK}.
Appearance of this logarithmic term also implies that the coordinate $r$ may
not be a good coordinate for the expansion of $B(r)$ in asymptotic region as
have been done in the global U(1) vortices \cite{Gre,HS}.

It is well-known that O(3) nonlinear $\sigma$ model in (2+1)D flat spacetime
supports self-dual solitons described by the first-order equation
\begin{eqnarray}\label{sdeq}
\partial_{i}\phi^{a}=\pm\frac{1}{v}\epsilon_{i}^{\;j}\varepsilon^{abc}\phi^{b}
\partial_{j}\phi^{c},
\end{eqnarray}
and any static regular topological soliton with finite energy satisfying
Euler-Lagrange equation is proved to be self-dual and to satisfy
Eq.~(\ref{sdeq}).
Here it would be natural to ask a question whether or not the obtained solutions
in anti-de Sitter space are self-dual. In curved spacetime,
 second-order equation
from the self-dual equation (\ref{sdeq}) is
\begin{eqnarray}\label{soeq}
\nabla^{2}\phi^{a}-\frac{1}{v^{2}}(\phi^{b}\nabla^{2}\phi^{b})\phi^{a}
=\pm\frac{1}{v}\varepsilon^{abc}(\partial_{j}\epsilon^{ji}+\Gamma^{j}_{jk}
\epsilon^{ki})\phi^{b}\partial_{i}\phi^{c},
\end{eqnarray}
where $\nabla^{2}$ denotes two-dimensional Laplacian.
In the static metric (\ref{cyl}), Eq.~(\ref{soeq}) becomes
\begin{eqnarray}\label{soeqr}
B\frac{d^{2}F}{dr^{2}} + \Bigl( B\frac{dN}{dr}
+\frac{dB}{dr} + \frac{B}{r} \Bigr)\frac{dF}{dr}
-\frac{n^2}{r^2}\sin F \cos F \nonumber \\
=\pm\frac{1}{v}\frac{e^N}{r}(B\frac{dN}{dr}+\frac{1}{2}\frac{dB}{dr})n\sin F.
\end{eqnarray}
Comparing Eq.~(\ref{soeqr}) with the Euler-Lagrange equation (\ref{eleq}),
we obtain a necessary condition for the metric, that is, the vanishment of the
right-hand side of Eq.~(\ref{soeqr}):
\begin{eqnarray}\label{condi}
\frac{dN}{dr}+\frac{1}{2B}\frac{dB}{dr}=0.
\end{eqnarray}
The solution of Eq.~(\ref{condi}) with a rescaling of time coordinate leads
to 
\begin{eqnarray}
ds^2=dt^2-dz^2-\frac{dr^2}{B(r)}-r^2d\theta^2 .
\end{eqnarray}
It is the very metric admitting
self-dual string-like solutions in curved spacetime with zero
cosmological constant~\cite{CG,GOR1}.
With the help of Eq.~(\ref{condi}), Eqs.~(\ref{Neq}) and (\ref{Beq}) are
reduced to an equation:
\begin{eqnarray}\label{fineq}
2|\Lambda|=-8\pi Gv^{2}\Big(\sqrt{B}\frac{dF}{dr}-\frac{n}{r}\sin F\Big)
\Big(\sqrt{B}\frac{dF}{dr}+\frac{n}{r}\sin F\Big).
\end{eqnarray}
Since the (anti-)self-dual solitons satisfying Eq.~(\ref{sdeq}) make the
right-hand side of Eq.~(\ref{fineq}) vanish,
we have $\Lambda=0$ as a necessary condition for any (anti-)self-dual soliton.
Therefore, the static
string-like topological and nontopological configurations of O(3) nonlinear 
$\sigma$ model under
the static metric (\ref{cyl}) cannot saturate Bogomolnyi-type bound in (anti-)de
Sitter spacetime. In fact static self-dual solitons of this model with
a cosmological constant was proved to be constructed only when the metric
is stationary and the cosmological constant is negative~\cite{KK}.

In this section we analyzed the O(3) nonlinear $\sigma$ model in anti-de Sitter
spacetime and found a new static soliton configuration whose nature is 
nontopological, and its topological charge is a multiple of half integer
in addition to the
well-known topological lump solution.
The obtained solitons are 
shown to be non-self-dual.

\section{Spacetime Structure}

We have obtained in the previous section all possible static regular soliton
solutions of Eq.~(\ref{eleq}), Eq.~(\ref{Neq}), and Eq.~(\ref{Beq}). 
In this section we address a question about possible spacetime 
manifolds formed by $\sigma$ soliton configurations and a negative vacuum
energy. Among the known (2+1)D anti-de Sitter spacetime solutions intriguing
ones are regular hyperboloid and BTZ black hole \cite{DJ,BTZ}.
In Ref.~\cite{KKK}, one of the authors showed that static global U(1) vortex can
form a space with two event horizons, which resembles a charged BTZ black hole.
Specifically,
what we are looking for is the existence of black hole horizon, which is
manifested by the region of non-positive $B(r)$.

At first let us investigate the structure of spatial manifolds by the
topological
lump solutions and show that any regular topological lump configuration
does not form
a BTZ-type black hole even when the magnitude of negative cosmological constant is small and the symmetry breaking scale is of the order of the Planck mass. 
{}From the asymptotic form of $B(r)$ in Eq.~(\ref{binf}),
one can easily read a necessary condition to have negative $B(r)$.
When $B_{\infty}$ is not negative, the series expansion (\ref{binf}) of $B(r)$  
is always positive for large $r$
and it implies impossibility of the existence of the horizon.
 On the other hand, Eq.~(\ref{bzero}) tells an
opposite possibility that $B(r)$ of an $n=1$ soliton can have zero at 
some $r$, if 
$4\pi G(B_0+n^2)F_0^2$ is much larger than the magnitude of cosmological 
constant $|\Lambda|$. In order to clarify this issue let us examine 
the integral equations for $N(r)$ and $B(r)$ obtained from Eq.~(\ref{Neq}) and 
Eq.~(\ref{Beq}):
\begin{eqnarray}
\label{Ninte}
N(r)&=&-8\pi G \int^\infty_r ds{}s\Big( \frac{dF}{ds}\Big)^2, \\
\label{Binte}
B(r)&=&e^{-N(r)}\Bigg\{ 2|\Lambda|\int^{r}_{0}{}ds{}s{}e^{N(s)}
-8\pi Gv^{2}n^{2}\int^r_0{}ds{}\frac{e^{N(s)}}{s}{}\sin^{2}F
+e^{N(0)}\Bigg\}.
\end{eqnarray}
First term in the square bracket of Eq.~(\ref{Binte}) describes 
contribution of the negative vacuum 
energy, 
and second term of it does the core mass.
In order to obtain negative $B(r)$ region for some $r$, 
small magnitude of the negative cosmological constant is favorable. 
Since the third term $e^{-N(0)}$ is of order one, another necessary
condition from the second term in Eq.~(\ref{Binte}) is 
the lower bound of symmetry breaking scale $v$ which must be
the Planck mass, i.e., $8\pi Gv^{2}\sim 1$.
To evaluate the value of $B_{\infty}$ in Eq.~(\ref{binf}), we take a crude
approximation such as
\begin{eqnarray}\label{Napp}
N(r)=0,
\end{eqnarray}
and
\begin{eqnarray}
F(r)= \left\{ \begin{array}{ll} 0 &\mbox{for $0< r< r_{c}-\Delta$}\\
\pi/2 &\mbox{for $r_{c}-\Delta\leq r\leq r_{c}+\Delta$}\\   
\pi &\mbox{for $r > r_{c}+\Delta$} \end{array}. \right.
\label{Fapp} 
\end{eqnarray}
Inserting Eqs.~(\ref{Napp}) and (\ref{Fapp}) into the integral equation
(\ref{Binte}) and comparing the result with Eq.~(\ref{binf}), we obtain
\begin{eqnarray}\label{Binfi}
B_{\infty}\sim 1-16\pi Gv^2n^2\Bigl( \frac{\Delta}{r_{c}}\Bigr).
\end{eqnarray}
Since both $r_{c}$ and $\Delta$ have the scale of soliton core size and then 
the ratio $\Delta/r_{c}$ is of the order one, we can confirm that the 
Planck scale as a symmetry breaking scale is necessary to exhibit the horizon 
of a BTZ black hole.

Now let us assume that there exists a horizon at $r_{H}$.
At each horizon a set of appropriate boundary conditions is 
\begin{eqnarray}
B(r_{H})&=&0,\\
\frac{dF}{dr}\bigg|_{r_{H}}&=&\frac{\frac{v^2n^2}{r^2_H}\sin 2F(r_{H})}{
16\pi Gr_H\Bigl(\frac{|\Lambda|}{4\pi G}-\frac{v^2n^2}{r^2_H}\sin^2F(r_{H}) 
\Bigr)}.
\label{fhori}
\end{eqnarray}
Since 
$B(0)=1$ and $B(r)\stackrel{r\rightarrow\infty}{\rightarrow}|\Lambda|r^{2}$, 
the region of negative $B(r)$ should be bounded and thereby the number of 
horizons should be even.
We attempt a series solution near the horizon $r_{H}$ to leading order:
\begin{eqnarray}
F(r)&\approx&F(r_{H})+\frac{\frac{v^2n^2}{r^2_H}\sin 2F(r_{H})}{
16\pi Gr_H\Bigl(\frac{|\Lambda|}{4\pi G}-
     \frac{v^2n^2}{r^2_H}\sin^2F(r_{H})\Bigr)}(r-r_H),
\label{horif}\\
N(r)&\approx&N(r_{H})+\frac{1}{32\pi Gr_H}\frac{
(\frac{v^2n^2}{r_H^2}\sin2F(r_{H}))^2}{\Bigl(\frac{|\Lambda|}{4\pi G}-
\frac{v^2n^2}{r_H^2}\sin^2F(r_{H}) \Bigr)^2}(r-r_H),\\
B(r)&\approx&8 \pi G r_H \Bigl(
\frac{ |\Lambda| }{ 4\pi G } - \frac{v^2n^2}{r_{H}^2}\sin^2F(r_{H}) \Bigr)
(r-r_{H}).
\label{horib}
\end{eqnarray}
Suppose that there exists a region of negative $B(r)$ bounded by $r_H^{in}$
and $r_H^{out}$ $(r_H^{in}<r<r_H^{out})$.
Then other necessary
conditions are $\frac{dB}{dr}\big|_{r_H^{in}}<0$ and
$\frac{dB}{dr}\big|_{r_H^{out}}>0$, and they lead to 
$\frac{|\Lambda|}{4\pi G}-\frac{v^2n^2}{(r^{in}_H)^{2}}\sin^2F(r^{in}_{H})<0$
and $\frac{|\Lambda|}{4\pi G}-\frac{v^2n^2}{(r^{out}_H)^{2}}
\sin^2F(r^{out}_{H})<0$ by Eq.~(\ref{horib}). However, now that $F(r)$ is 
monotonically increasing from $F(0)=0$ to $F(\infty)=\pi$, 
the negativity of the numerator of the second term in Eq.~(\ref{horif})
forces a condition to $F(r)$, that
the value of $F(r_H^{in})$ should be larger than $\pi/2$ and that of 
$F(r_H^{out})$ should be smaller than $\pi/2$. 
Therefore above conclusion, i.e., $F(r_H^{in})>F(r_H^{out})$, contradicts
to the monotonically increasing property of $F(r)$.
Therefore we arrive at a no-go conclusion that axially symmetric
regular static topological lump solution in O(3) nonlinear 
$\sigma$ model cannot 
support a BTZ-type black hole with two horizons in anti-de Sitter spacetime.

Since we have proved that any $B(r)$ corresponding to regular topological lump
 configuration
cannot be negative, the remaining question for the 
nonexistence
of the black hole horizon is to show the positivity of the minimum of $B(r)$.
Again, let us assume that there exists a point $r_H$ such that $B(r_H)=0$
and this is the minimum value of $B$. Then the 
position of the horizon $r_{H}$ and the value of $F(r_{H})$ are determined in
a closed form from Eqs.~(\ref{eleq}) and ~(\ref{Beq}):
\begin{eqnarray}\label{extho}
r_{H}=\sqrt{\frac{4\pi Gv^2n^2}{|\Lambda|}} ~~~~{\rm and}~~~~~
F(r_{H})=\frac{\pi}{2}.
\end{eqnarray}
If there exists regular solution to have $B(r_{H})=0$, one can try a series
expansion around the horizon $r_{H}$ such as
\begin{eqnarray}
\label{extf}
F(r)&\approx&\frac{\pi}{2}+f_{1}(r-r_{H})+f_{2}(r-r_{H})^{2}+f_{3}(r-r_{H})^{3}
+\cdots, \\
\label{extb}
B(r)&\approx&B_{2}(r-r_{H})^{2}+B_{3}(r-r_{H})^{3}+\cdots .
\end{eqnarray}
After replacing $N(r)$ dependent term in Eq.~(\ref{eleq}) by use of 
Eq.~(\ref{Neq}), we substitute Eq.~(\ref{extf}) and Eq.~(\ref{extb}) into the
modified equations (\ref{eleq}) and (\ref{Beq}). The comparison of both sides 
of the equations results in the trivial solution of $F(r)$, i.e.,
$0=f_{1}=f_{2}=f_{3}=\cdots$. It means that the topological lump which is 
a nontrivial solution cannot constitute spatial manifold of an extremal 
black hole with one horizon. Combining with the previous proof, we conclude that
any regular topological lump of O(3) nonlinear $\sigma$ model does not form  
spacetime of a BTZ black hole irrespective of the values of $|\Lambda|/v^{2}$
and $8\pi Gv^{2}$. Therefore, the shapes of $B(r)$ from the regular 
topological lump solutions are 
classified into two categories: One is monotonically increasing $B(r)$
and the other is convex down $B(r)$ (See Fig.~\ref{fig5}).

\begin{figure}

\setlength{\unitlength}{0.1bp}
\begin{picture}(3600,2160)(0,0)
\put(1950,150){\makebox(0,0){{\Large $vr$}}}
\put(100,1230){%
\makebox(0,0)[b]{\shortstack{{\Large $B$}}}%
}
\put(3550,300){\makebox(0,0){10}}
\put(2910,300){\makebox(0,0){8}}
\put(2270,300){\makebox(0,0){6}}
\put(1630,300){\makebox(0,0){4}}
\put(990,300){\makebox(0,0){2}}
\put(350,300){\makebox(0,0){0}}
\put(300,2060){\makebox(0,0)[r]{5}}
\put(300,1728){\makebox(0,0)[r]{4}}
\put(300,1396){\makebox(0,0)[r]{3}}
\put(300,1064){\makebox(0,0)[r]{2}}
\put(300,732){\makebox(0,0)[r]{1}}
\put(300,400){\makebox(0,0)[r]{0}}
\end{picture}

\begin{center}{(a)}
\end{center}


\setlength{\unitlength}{0.1bp}
\begin{picture}(3600,2160)(0,0)
\put(2000,150){\makebox(0,0){{\Large $\ln vr$}}}
\put(100,1230){%
\makebox(0,0)[b]{\shortstack{{\Large $B$}}}%
}
\put(3268,300){\makebox(0,0){6}}
\put(2705,300){\makebox(0,0){4}}
\put(2141,300){\makebox(0,0){2}}
\put(1577,300){\makebox(0,0){0}}
\put(1014,300){\makebox(0,0){-2}}
\put(450,300){\makebox(0,0){-4}}
\put(400,1783){\makebox(0,0)[r]{1.5}}
\put(400,1322){\makebox(0,0)[r]{1}}
\put(400,861){\makebox(0,0)[r]{0.5}}
\put(400,400){\makebox(0,0)[r]{0}}
\end{picture}

\begin{center}{(b)}
\end{center}

\vspace{3mm}

\caption{Two characteristic shapes of $B(r)$ formed by the topological
lumps: (a) A monotonically increasing $B(r)$ when $8\pi Gv^{2}=8\times 
10^{-8}$, 
 $|\Lambda|/v^{2}=0.04$, and $F_{0}=1250$, 
(b) A convex down $B(r)$ when $8\pi Gv^{2}=0.2$, 
$|\Lambda|/v^{2}= 4.0\times 10^{-6}$, and $F_0=5.896$.}
\label{fig5}
\end{figure}

Behavior of $B(r)$ given in Fig.~\ref{fig5} describes the structure of
the spatial hypersurface of the (2+1)-dimensional spacetime. 
Since the metric is static,
spatial manifold is characterized by the circumference $l(r)\equiv 2\pi r$
and the radial distance ${\cal R}(r)=\int_{0}^{r}dr/\sqrt{B(r)}$.  
We embed it into a three-dimensional hyperbolic 
space by introducing the third axis $Z$ such that ${\cal R}^{2}=-Z^{2}
+r^{2}/B_{m}$, where $Z\geq 0$ and $B_m$ is the minimum of $B(r)$.  
For sufficiently large $r$, 
$B(r)\sim|\Lambda|r^{2}+B_{\infty}$ as given in Eq.~(\ref{binf}).
Introducing variables such as $\sqrt{|\Lambda|/B_{\infty}}\,\,r=\sinh\chi$ and 
$\sqrt{B_{\infty}}\,\,\theta=\Theta$, we obtain the asymptotic metric
\begin{eqnarray}
ds^2\approx \frac{1}{|\Lambda|}(d\chi^2+\sinh^2\chi d\Theta^2).\label{hyperm}
\end{eqnarray}
The asymptotic region of two-dimensional spatial manifold given by 
Eq.~(\ref{hyperm}) is a hyperboloid 
with deficit angle $2\pi(1-\sqrt{B_{\infty}}\,)$. 
By use of Eq.~(\ref{Binfi}) we estimate the deficit angle to be $16\pi^{2}
Gv^{2}n^{2}$.
This can easily be understood by the 
nonexistence of a long tail term in energy-momentum tensor. Since 
nonvanishing independent components of it are
\begin{eqnarray} 
T^t{}_t&=&\frac{v^{2}}{2}B\Bigl({\frac{dF}{dr}}\Bigr)^2+
             \frac{n^2v^{2}}{2r^2}\sin^2F,
\label{ttt}\\
T^r{}_r&=&-\frac{v^{2}}{2}B\Bigl({\frac{dF}{dr}}\Bigr)^2+
          \frac{n^2 v^{2}}{2r^2}\sin^2F,
\label{trr}
\end{eqnarray}
they look to include a long tail term. 
However, substituting Eq.~(\ref{finf}) into
Eqs.~(\ref{ttt}) and (\ref{trr}), we read that the leading term is
${\cal O}(1/r^{4})$ term which does not affect the asymptotic 
region of two-dimensional
spatial manifold. 

As we can expect from Fig.~\ref{fig5}, the spatial manifold 
on the core of topological lump is involved in one of two categories. When
the absolute value of negative cosmological constant is large enough, i.e.,
$|\Lambda|/v^{2}>8\pi GF_{0}^{2}\delta_{1n}$ and $B_{m}=1$, 
the relation between $Z$ and
$r$ near the origin is 
$dZ\approx\sqrt{\alpha r^2/(1+\alpha r^2)}\,\,dr$
where $\alpha\equiv |\Lambda|-8\pi Gv^{2}F_{0}^{2}\delta_{1n}$.
Then the core region of this soliton is also hyperbolic,
$(Z+1/\sqrt{\alpha})^2 - r^2=1/\alpha$. 
On the other hand, when $B(r)$ is decreasing near the origin, i.e., 
$|\Lambda|/v^{2}<8\pi GF_{0}^{2}\delta_{1n}$ 
and $0<B_{m}<1$, the relation between
$Z$ and $r^{\prime}(\equiv r/\sqrt{B_{m}})$ is given in the following:
\begin{eqnarray}\label{Fig6}
Z(r)\approx \left\{ \begin{array}{ll}
\displaystyle{\sqrt{1-B_m}\,r^{\prime}\Bigl(1+
\frac{\alpha r^{\prime 2}}{ 6(1-B_m)}\Bigr)} 
 & \mbox{for small} \,\,r^{\prime}\\
\displaystyle{\sqrt{\frac{B_{\infty}}{|\Lambda|B_m}+r^{\prime 2}}\,\,
-\sqrt{\frac{B_{\infty}}{|\Lambda|B_m}}} &\mbox{for large} \,\,r^{\prime},
\end{array}
\right.
\end{eqnarray} 
and
\begin{eqnarray}\label{infle}
dZ\approx \sqrt{B_{m2}}(r^{\prime}-r^{\prime}_m)dr^{\prime}
~~~~\mbox{around}\,\, r^{\prime}=r^{\prime}_m(\equiv r_m/\sqrt{B_m}\, ),
\end{eqnarray}
where $B_{m2}$ is the coefficient of the second order term in the series
of $B(r)$ around $r_m$.
Since $\alpha$ is negative, the first line in Eq.~(\ref{Fig6}) tells us that
the core region is convex up. In order to
connect smoothly the core and asymptotic regions of the spatial manifold,
there should exist an inflection point about the minimum point $r_{m}$ of 
$B(r)$
as given in Eq.~(\ref{infle}).

{}From now on let us look into possible structure of a spacetime manifold
formed by the nontopological soliton of half integral winding. 
Recalling the asymptotic form
of $B(r)$ in Eq.~(\ref{balpha}), one may easily notice a difference between
this equation and Eq.~(\ref{binf}) for the topological lump:
The asymptotic space of the half integral winding soliton includes a logarithmic
term with negative coefficient. This metric function is 
the same as that of a global U(1) vortex \cite{KKK}. In the model 
of a complex scalar field the very
logarithmic term has played a crucial role to constitute a vortex 
 BTZ black hole with two horizons. 
On the other hand, our nontopological $\sigma$ solitons are distinguished 
from global U(1) vortices by the following points. For a given model with
fixed model parameters, global U(1) vortex solution is unique, however, there
are many nontopological $\sigma$ soliton solutions characterized
by the maximum value of scalar amplitude which is larger than $\pi/2$ but 
smaller than $\pi$. About the shape of scalar amplitude, the former is a
monotonically increasing function from zero to the vacuum expectation value 
but the latter can contain oscillatory behavior as shown in 
Fig.~\ref{fig4}. 
Therefore, nontopological $\sigma$ solitons with the same topological charge
are classified into many subclasses by the number of nodes.

\begin{figure}

\setlength{\unitlength}{0.1bp}
\begin{picture}(3600,3240)(0,0)
\put(1700,2840){\makebox(0,0){(a)}}
\put(2200,1100){\makebox(0,0){(b)}}
\put(2600,850){\makebox(0,0){(c)}}
\put(3000,800){\makebox(0,0){(d)}}
\put(2000,150){\makebox(0,0){{\Large $vr$}}}
\put(100,1770){%
\makebox(0,0)[b]{\shortstack{{\Large $B$}}}%
}
\put(3550,300){\makebox(0,0){10}}
\put(2930,300){\makebox(0,0){8}}
\put(2310,300){\makebox(0,0){6}}
\put(1690,300){\makebox(0,0){4}}
\put(1070,300){\makebox(0,0){2}}
\put(450,300){\makebox(0,0){0}}
\put(400,3140){\makebox(0,0)[r]{1.2}}
\put(400,2683){\makebox(0,0)[r]{1}}
\put(400,2227){\makebox(0,0)[r]{0.8}}
\put(400,1770){\makebox(0,0)[r]{0.6}}
\put(400,1313){\makebox(0,0)[r]{0.4}}
\put(400,857){\makebox(0,0)[r]{0.2}}
\put(400,400){\makebox(0,0)[r]{0}}
\end{picture}

\vspace{3mm}

\caption{Plots of $B(r)$ for various $F_0$'s for $|\Lambda|/v^{2}=0.01$
and $8\pi Gv^{2}=0.4$: (a) zero node, (b) one node, (c) two nodes,
(d) extremal ($F_0 = 2.41902$ up to $10^{-6}$ precision).}
\label{fig6}
\end{figure}
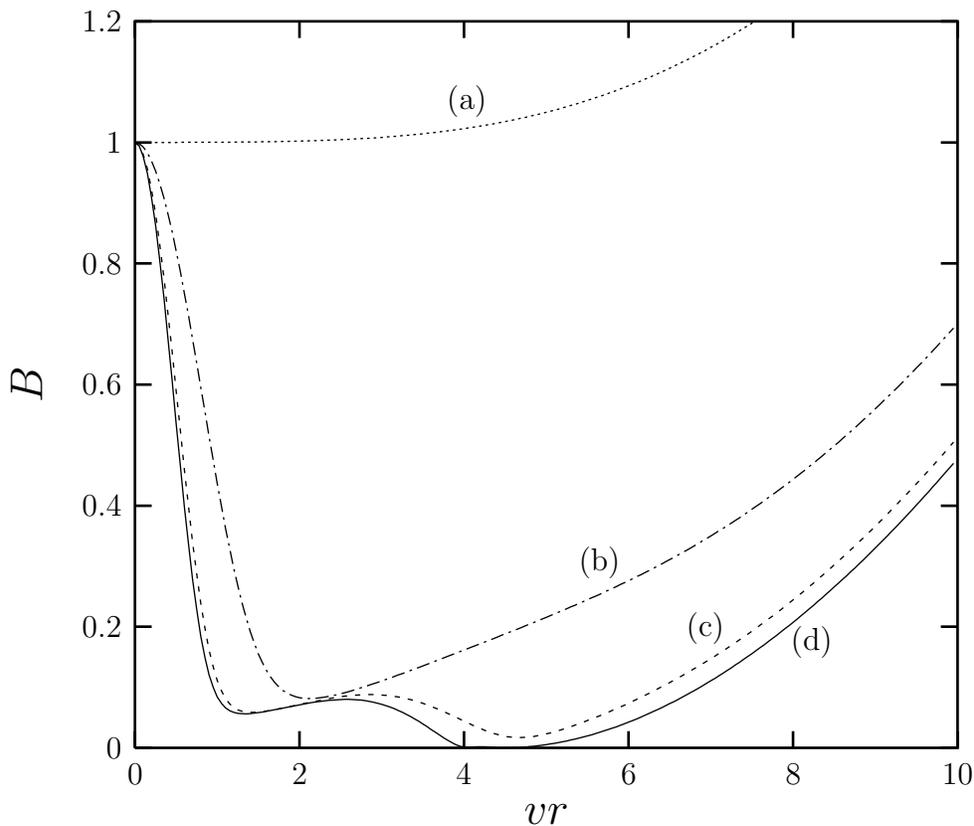

The existence of the logarithmic term in the asymptotic form (\ref{balpha}) of
the metric function $B(r)$ lets us ask an intriguing question about the
generation of BTZ black hole for a small magnitude of cosmological constant
and relatively large symmetry breaking scale as happened in gravitating global
U(1) vortices with a negative cosmological constant. The results of the
numerical analysis are summarized in Figs.~\ref{fig4} and~\ref{fig6}. 
Figure 6 shows the metric
$B$ as a function of $r$ for various number of nodes. As the number of nodes
increases (or equivalently the value of $F_0$ in Eq.~(\ref{fzero}) increases),
the value of the minimum of $B$ decreases. It is also natural that the
behavior of $B$ is as like as Fig.~\ref{fig6} as the symmetry breaking scale is
increased with a fixed value of $F_0$. The nontopological $\sigma$ soliton 
solutions 
are seen to tend towards black hole solutions as the symmetry breaking scale
$v$ or the number of nodes is increased, as might be expected. A difference
from the behavior of $B$ for global U(1) vortices can be noticed: 
In case of
the global U(1) vortices, 
one bump was dug and such minimum of $B$ finally touched
zero value \cite{KKK}, however several bumps are developed for nontopological
$\sigma$ solitons and the outmost one becomes the minimum of $B$ and then this
position tends to be a horizon as shown in Fig.~\ref{fig6}. The graphs in 
Fig.~\ref{fig4}
show that wiggles of the scalar field tend to subside to the boundary value
$\pi/2$ outside the location of the minimum of $B$. 

Within our numerical precision, a careful analysis of solutions near the
transition to a black hole indicates that the nontopological $\sigma$ soliton
looses its scalar amplitude hair as it develops a horizon. In fact, it is
predictable from Eq.~(\ref{horif}): When $F(r_{H})=\pi/2$, the actual value of
$dF/dr|_{r_{H}}$ vanishes for any extremal black hole. Here let us write down
the action (\ref{act}) in terms of stereographically projected variables, i.e.,
$\phi^{a}=v(\sin F\cos(\Theta+\eta), \sin F\cos(\Theta+\eta), \cos F)$, where
the multi-valued $\Theta$ represents the topological sector and the
single-valued function $\eta$ does the 
Goldstone degree for a given topological
sector. Then, in (2+1)D flat spacetime, we obtain
\begin{eqnarray}\label{fthe}
{\cal L}=\frac{v^{2}}{2}\Big[\partial_{\mu}F\partial^{\mu}F+
\sin^{2}F\partial_{\mu}(\Theta+\eta)\partial^{\mu}(\Theta+\eta)\Big].
\end{eqnarray}
By use of a duality transformation in 2+1 dimensions \cite{DS}, one can easily
show in the context of the path integral formulation that the above theory
(\ref{fthe}) is equivalent to that of a U(1) vector field $A_{\mu}$:
\begin{eqnarray}\label{dthe}
{\cal L}=\frac{v^{2}}{2}\partial_{\mu}F\partial^{\mu}F-
\frac{1}{4}\frac{F_{\mu\nu}F^{\mu\nu}}{\sin^{2}F}+\frac{v}{2}
\epsilon^{\mu\nu\rho}F_{\mu\nu}\partial_{\rho}\Theta ,
\end{eqnarray}
where $F_{\mu\nu}=\partial_{\mu}A_{\nu}-\partial_{\nu}A_{\mu}$. 
If the scalar amplitude is frozen to be $F=\pi/2$, outside the black hole
horizon, then the matter field action
(\ref{dthe}) is nothing but the sum of the Maxwell term and the minimal
interaction between the gauge field and point particle.
Now we understand the reason why a $\sigma$ soliton
black hole looks just like a charged BTZ black hole outside 
the horizon\cite{BTZ}.
Therefore, nontopological $\sigma$ solitons in O(3) nonlinear $\sigma$ model 
do not break no-hair theorem.
This phenomenon seems universal for our nontopological soliton 
solutions since it happens for a wide
range of the symmetry breaking scale $v$ and the negative cosmological
constant $\Lambda$. In this aspect, the regular nontopological $\sigma$
solitons are also distinctive from the topological global U(1) vortices with
scalar hair \cite{KKK}, but resemble the case of regular gravitating
magnetic monopoles in 3+1 dimensions \cite{LNW}. 
We can imitate the case of exact singular monopole solution 
whose metric is the
Reissner-Nordstr\"{o}m black hole \cite{BR}. Specifically, $F(r)=\pi/2$,
$\Theta = n\theta $, and $\eta = 0$, 
everywhere and the corresponding black hole spacetime is a charged BTZ-type.
More plausible singular configurations may be obtained by changing the boundary condition of the metric function at the origin, i.e., $B(0)\neq 1$ similar to 
the monopole black hole\cite{LNW}.
Since the singularity
of the fields which is presumably at the origin can be hidden behind a 
horizon, we may not exclude the possibility that singular 
solutions can form small BTZ black holes lying within a nontopological $\sigma$
soliton. Since no non-Abelian scalar hair can penetrate 
the horizon for regular solitons, 
we can evaluate the
position of the horizon by using Eqs.~(\ref{eleq}) and (\ref{Beq}), 
and it is nothing 
but the
formula (\ref{extho}). The values of the horizon obtained by numerical
analysis coincide with those from Eq.~(\ref{extho}) within precision.

As mentioned previously,
we have many nontopological soliton excitations classified by 
the number of nodes for a given topological sector of the 
theory so that we have to discuss
stability among these classical solutions carrying
with the same topological charge. 
A good method
is to compare their masses.
Since the obtained spacetime is not asymptotically flat but is hyperbolic,
the usual Arnowitt-Deser-Misner mass is not obtained in the limit $r
\rightarrow\infty$. For the energy per unit length of infinitely-long
axially symmetric systems, known expressions are the C-energy \cite{Tho} 
and the
conserved quasilocal mass \cite{BY}. Here we use the latter of which
expression for the static observer is given by
\begin{eqnarray}\label{defm}
M_{q}&\equiv& \frac{1}{4G}\sqrt{e^{2N(r)}B(r)}\Big(\sqrt{|\Lambda|r^{2}+1}
-\sqrt{B(r)}\Big),\\
&\stackrel{r\rightarrow\infty}{\longrightarrow}& \left\{
\begin{array}{ll} 
\displaystyle{2\pi n^2v^2\Bigl(\frac{\Delta}{r_{core}}}\Bigr)
 &\mbox{for the topological lump}\\
\displaystyle{\pi n^2v^2\Bigl[\ln\Bigl(\frac{r}{r_{core}}\Bigr)+
2\sin^2\beta\Bigl(\frac{\Delta}{r_{core}}\Bigr)\Bigr] }
&\mbox{for the nontopological soliton},
\end{array}
\right.
\end{eqnarray}
where Eq.~(\ref{binf}) was used in the right-hand side of the above expression.
The mass for the topological lump has only the constant term. It is obvious
because this lump is localized around its core without a long range tail
term. The nontopological soliton of half integral winding has 
logarithmically divergent 
mass term in 
addition to the core mass. It shows some resemblance between static global
U(1) vortex and the nontopological soliton in O(3) nonlinear $\sigma$
model, whose leading long range term is the same, i.e., $T^{t}_{\;t}
\sim 1/r^{2}$ for large $r$.

For the $n=1$ class of solutions we compare the values of the quasilocal mass 
(\ref{defm}) of no node solution, one node solution, two node solution, the
solution of an extremal black hole, charged BTZ black hole at a sufficiently 
large distance $vr=50$ as a function of $v$ with fixed $8\pi G=0.4$ 
and $|\Lambda|=0.01$ (See Table 1).
The tendency that the quasilocal mass increases for higher node solutions
looks
universal, and further numerical studies for various $G$, $|\Lambda|$, and
$v$ also keep the same behavior. Therefore, no node solution is the lowest
energy solitonic excitation among those with a given charge $n/2$. Since
(2+1)D Einstein gravity does not have any attractive 
propagating gravitational degree,
it seems natural. All half integral winding solitons are nontopological, so
excited spectra may decay into the no node soliton of the lowest energy. 
This procedure
may presumably be correct for the solitons in the space of a regular
hyperboloid because the system has massless Goldstone degrees. Now, if we
recall that no node solution with the monotonically increasing $F(r)$ cannot
form a black hole horizon, then an intriguing question is raised about the
stability of an extremal BTZ-type black hole. In 3+1 dimensions, attractive
gravitational force usually makes a matter distribution with mass larger than
the critical value unstable and leads to the gravitational collapse where
the destination is the formation of a black hole. It seems unlikely for our
O(3) nonlinear $\sigma$ model in (2+1)D anti-de Sitter spacetime. On the other
hand, there may be an opposite procedure that an extremal BTZ-type black hole
is produced but it is energetically unfavorable 
and then the horizon disappears.
However, we need further study on the stability of nontopological solitons to
settle down this issue.
Now a comment about the critical symmetry breaking scale
is in order. In any natural environment the magnitude of negative cosmological 
constant is
much lower than the symmetry breaking scale $v$, and the very symmetry
breaking scale $v$ is
much lower than the Plank scale. For example, if we consider a present
universe with an extremely small bound of the negative cosmological constant
$(|\Lambda|\sim 10^{-83}{\rm Gev}^2)$, the critical value of the symmetry
breaking not to form a BTZ-type black string
is about $10^{-2}$eV which is very low energy. Of course, the above 
estimation is far from realistic
situation before we take into account the anisotropy in cosmic ray background
and other cosmological fluctuations.

\begin{center}{
\begin{tabular}{|c|c|c|c|c|} \hline
node    & 0       & 1       &   2     & extremal  \\ \hline
$v=1$   & 0.01245 & 0.01918 & 0.02056 & 0.02080   \\ \hline
$v=1.5$ & 0.02144 & 0.02625 & 0.02646 & 0.02647   \\ \hline
$v=2$   & 0.02664 & 0.02840 & 0.02841 & 0.02841   \\ \hline
\end{tabular}
}
\end{center}

\vspace{5mm}

Table 1. The values of quasilocal mass of various node solutions and the
extremal charged BTZ black hole at a large distance $vr=50$ with $8\pi G=0.4$
and $|\Lambda|=0.01$.

\section{Geodesic Motions}

The study of time-like and null geodesics is an adequate way to visualize 
the form of interaction on the soliton and the feature of its spacetime.
Let us analyze possible geodesic motions and clarify whether a
test particle experiences attraction or repulsion due to the soliton. The 
geometry depicted by Eq.~(\ref{cyl}) admits the rotational killing vector
$\partial/\partial\theta$ and the static killing vector 
$\partial/\partial t$, so two corresponding constants of motion along
geodesics are
\begin{eqnarray}\label{cofm}
\gamma=Be^{2N}\frac{dt}{ds}\;\;\mbox{and}\;\;L=r^{2}\frac{d\theta}{ds},
\end{eqnarray}
where $s$ is an affine parameter along the geodesic. Since the space is not 
asymptotically flat, the constant $\gamma$ cannot be interpreted as the local
energy of the test particle at infinity. The radial geodesic equation is
\begin{eqnarray}
\label{radial}
\frac{1}{2}\Big(\frac{dr}{ds}\Big)^2&=& -\frac{1}{2}\bigg[
B(r)\Big(m^2 +\frac{L^2}{r^2}\Big)
-\frac{\gamma^2}{e^{2N(r)}}\bigg]=-V(r),
\end{eqnarray}
where the mass of the test particle can be rescaled as $m=1$ for time-like
geodesics and $m=0$ for null geodesics.
We analyze the trajectories of test particles for the topological lump
background and the nontopological soliton background separately, and they are 
divided into four categories according to whether they have mass $( m=1 )$ or
not $( m=0 )$, or whether their motions are purely radial $(L=0)$ or rotating
$(L\neq 0)$. As shown in 
Fig.~\ref{fig5} and Fig.~\ref{fig6}, the geometry of spatial manifolds of our 
$\sigma$ model solitons is similar to those of global U(1) vortices \cite{KKK}.
Here we briefly mention different points.

\subsection*{\bf {IV.1 \,\,\,  Topological Soliton }}

The main character of the spacetime structure of topological lumps is the
absence of black hole. Due to this character, the geodesic motions are 
simple. It is qualitatively similar to the regular hyperboloids by global U(1) 
vortices \cite{KKK}.  

For the radial motion $(L=0)$ of a massless test particle $(m=0)$,
$B(r)$ dependence disappears in the effective potential $V(r)$.
The allowed motions are (i) stopped particle motion for $\gamma=0$ and 
(ii) unbounded motion for $\gamma\neq 0$ with the speed 
$dr/ds=\gamma/\sqrt{2}$ at spatial infinity. 
Since $N(r)$ is monotonically increasing,
this massless test particle in a radial motion always feels attractive force.

For the rotational motions $(L\neq 0)$ of a massless
test particle $(m=0)$, the effective potential 
includes the centrifugal force term $L^{2}B(r)/2r^{2}$ which forbids the test
particle to access the soliton core. Therefore, any allowed rotational
motion should have the minimum value of radius $r_{min}$ that $r\geq r_{min}$.
Since the value of the effective potential is 
$(|\Lambda|L^{2}-\gamma^{2})/2$ at spatial infinity, any allowed motion 
should be bounded by the minimum radius $r_{min}$ and the maximum radius 
$r_{max}$ when $|\Lambda|L^{2}>\gamma^{2}$.
However we cannot see this easily
due to the smallness of $|\Lambda|$. When $|\Lambda|L^{2}\leq\gamma^{2}$,
the motions are also divided into two classes by the peak speed: One is
the class with the peak speed at infinity,
and the other is that with the peak speed at a finite radius.

The effective potential for the radial motion $(L=0)$
of a massive test particle $(m=1)$ is
\begin{eqnarray}\label{m1l0pot}
V(r)=\frac{1}{2}\bigg(B(r)-\frac{\gamma^{2}}{e^{2N(r)}}\bigg).
\end{eqnarray}
For large $r$, it is approximated as
\begin{eqnarray}
V(r)\approx \frac{|\Lambda|}{2}r^{2}+\frac{1}{2}(B_{\infty}-\gamma^{2})
+{\cal O}(1/r^{2}),
\end{eqnarray}
and then all possible motions are bounded. Since the power series expansion
of $V(r)$ for small $r$ is
\begin{eqnarray}\label{m1l0r0}
V(r)\approx \frac{1}{2}(1-\gamma^{2}e^{-2N(0)})+\bigg[\Big(\frac{1}{2}
|\Lambda|-
4\pi Gv^2F_0^2(1-\gamma^2 e^{-2N(0)})\Big) r^2\bigg]+\cdots,
\end{eqnarray}
we divide the shapes of the potential (\ref{m1l0pot}) into two classes.
When the negative vacuum energy dominates the repulsive force of the 
scalar field
even at the core of the soliton, i.e., $|\Lambda|/2-
 4\pi Gv^2F_0^2(1-\gamma^2 e^{-2N(0)})\geq 0$, $V(r)$ is monotonically
increasing and thereby the force is attractive everywhere.
Then the minimum of the effective potential is at the origin and its value is
$|\Lambda|/16\pi Gv^2F_0^2$. 
The leading constant term in Eq.~(\ref{m1l0pot}), which is the minimum of 
$V(r)$, tells us that the radial motions are allowed only when $\gamma\geq
e^{N(0)}$. On the other hand, when $|\Lambda|/2-
 4\pi Gv^2F_0^2(1-\gamma^2 e^{-2N(0)})\leq 0$, the test particle
with $\gamma$ smaller than the critical value $\gamma_{cr}$ $(\gamma_{cr}
= e^{N(0)}\sqrt{1-|\Lambda|/16\pi Gv^2F_0^2}~)$ feels repulsive force at the 
core of the soliton. The allowed value of
$V(0)$ lies between $|\Lambda|/16\pi Gv^2F_0^2$ and $1/2$. One may expect
that there exists the negative region of $V(r)$ between $r_{min}$ and
$r_{max}$, however 
our numerical work shows the absence of such region. 
Possible motions are (i) the stopped motion, (ii) the oscillation between 
the minimum radius and the maximum radius, (iii) rolling to the origin, as 
$\gamma$ decreases.

For the circular motions $(L\neq 0)$ of a massive test
particle $(m=1)$, the effective potential takes general form 
(See Eq.~(\ref{radial})). 
Since the centrifugal force term dominates at small $r$,
$V(r)$ for small $r$ resembles that of the case of a rotating motion of a 
massless test particle, and there exists perihelion
$r_{min}$. For large $r$, all motions are bounded by an aphelion $r_{max}$
because of the negative cosmological constant term. The allowed motions are 
(i) the circular orbit at $r_{circ}$ when $\gamma=\gamma_{circ}$, 
and (ii) the bounded orbit between perihelion $r_{min}$ and
aphelion $r_{max}$ when $\gamma$ is larger than $\gamma_{circ}$.
Noticing the vanishment of $V(r)$ at both $r_{min}$ and $r_{max}$, one may 
suspect that the comoving time defined by
\begin{eqnarray}\label{timeho}
\tau=\int dr\frac{\gamma}{\sqrt{-2V(r)}},
\end{eqnarray}
diverges when the test particle approaches to those points. However, since
the denominator in Eq.~(\ref{timeho}) is proportional to 
$1/\sqrt{r-r_{min}}$
(or $1/\sqrt{r-r_{max}}$), it takes finite comoving time to reach a boundary. 
So does the coordinate time defined by $dt/d\tau=\gamma/Be^{-2N}$ since
there is no black hole horizon, i.e., $B(r)>0$ for all $r$.

\subsection*{\bf {II.2\,\,\, Nontopological Soliton }}
 
 As we have discussed in the previous section, there exist black hole 
solutions for some nontopological solitons. 
For some regular solutions, e.g., (a) and (b) in 
 Fig.~\ref{fig6}, the geodesic motions are not so much different 
from those of 
 topological solitons. 
There are different $B$'s with several bumps as shown in the graphs 
(c) and (d) in Fig.~\ref{fig6}. One may suspect that these $B$'s generate
different geodesic motions, e.g., two isolated radial regions in the effective 
potential $V(r)$.
However, our numerical works show that there 
are no such effective potential, so that the character of geodesic motions for 
regular nontopological solitons is the same as that for topological lumps.  
The only difference is the rapid variation of $V(r)$ near the origin, due to 
rapidly increasing $N(r)$.
Note that Eq.~(\ref{Neq}) reflects the rapid increasing of $N(r)$ for 
many nodes of our nontopological soliton.

{}From Eqs.~(\ref{cofm}) and (\ref{radial}),
the elapsed coordinate time $t$ of a test particle
which moves from $r_0$ to $r$ is
\begin{eqnarray}
t=\int^{r_0}_{r}\frac{dr}{B(r)e^{N(r)}
\sqrt{1-\frac{1}{\gamma^2}(m^{2}+\frac{L^2}{r^{2}})
B(r)e^{2N(r)}}}.
\end{eqnarray}
It diverges when the test particle approaches to a point where $B(r)$ vanishes 
at least linearly.
As we  expected,
the spacetime with horizons depicts that of a black hole.
For the black hole solutions, 
our geodesic motions outside the horizon are intrinsically the same with 
that of a charged BTZ black hole, since any scalar hair does not penetrate 
the horizon but the logarithmic Goldstone sector. 

As usual, the matter distribution is reflected to the scalar curvature which
is given by
\begin{eqnarray}\label{scurv}
R=-6\Lambda-16\pi GT^{\mu}_{\;\mu}.
\end{eqnarray}
For small $r$, Eq.~(\ref{scurv}) for both the topological
lump and the nontopological soliton becomes
\begin{eqnarray}
R\approx 6|\Lambda|-8\pi Gn^2v^{2}F^{2}_{0}[2+(|\Lambda|-
8\pi Gv^2F_{0}^2\delta_{1,n})r^2]r^{2n-2}.
\end{eqnarray}
When $n=1$, the curvature can be negative due to the accumulation of the matter
at the core of the soliton at the Planck scale. 
For large $r$, the behavior of the scalar curvature
depends on the characteristic of the solitons:
\begin{eqnarray}
R\approx
\left\{\begin{array}{ll}
\displaystyle{6|\Lambda|-32\pi Gv^{2}F^{2}_{\infty}|\Lambda|\frac{1}{r^{4}}} &
\mbox{for the topological lump} \\
\displaystyle{6|\Lambda|-8\pi Gv^2n^2\frac{1}{r^2}
-32\pi Gv^2|\Lambda|F^2_{\pi/2,\infty}\frac{1}{r^4} }
& \mbox{for the nontopological soliton}.
\end{array}\right.
\end{eqnarray}
As expected, the space is curved at large $r$ for the nontopological soliton,
while it is not for the topological lump.
Although we have charged BTZ black holes from some half 
integral winding soliton configurations,
we may expect that all the obtained spacetimes do not contain physical
curvature singularity due to the regularity of the matter fields and the
metric functions everywhere. It is easily checked by the Kretschmann scalar,
\begin{eqnarray}
\lefteqn{R_{\mu\nu\rho\sigma}R^{\mu\nu\rho\sigma}
= 4G_{\mu\nu}G^{\mu\nu}}\label{curv}\\
&=& 4\mbox{Tr}
\bigg[\mbox{diag}\Big(-\frac{1}{2r}\frac{dB}{dr},-\frac{1}{2r}\frac{dB}{dr}
-\frac{B}{r}\frac{dN}{dr}, -\frac{1}{2}\frac{d^{2}B}{dr^{2}}-\frac{3}{2}
\frac{dB}{dr}\frac{dN}{dr}-B\frac{d^{2}N}{dr^{2}}-B\Big(\frac{dN}{dr}\Big)^{2}
\Big)\bigg].\nonumber 
\end{eqnarray}
When both $N(r)$ and $B(r)$ are regular everywhere, the only possible 
singularity can be at the origin in Eq.~(\ref{curv}), however it is also 
regular at the origin due to the behaviors of those metric functions
at the origin as given in Eqs.~(\ref{nzero}) and (\ref{bzero}). Then,
the spacetime formed by the topological lump or the nontopological soliton is
always regular everywhere irrespective of the existence of the black hole
horizon.

\section{Conclusion and discussion}

In this paper we have studied static soliton solutions of O(3) nonlinear
$\sigma$ model coupled to Einstein gravity with a negative cosmological
constant. It has been shown that any regular static soliton configuration
with axially symmetric static metric is not self-dual in this anti-de 
Sitter spacetime. By examining second order Euler-Lagrange equations, we
obtained a new class of nontopological soliton solutions whose winding
number is multiple of half integer in addition to the well-known topological
lumps with integral topological charge. Scalar amplitude of the 
topological lump solution is monotonically increasing function which
interpolates the symmetric vacuum and the broken vacua, and its energy
density, the time-time component of energy-momentum tenser, is 
localized around
the soliton core. The lack of a long tail term in the energy density at 
asymptotic region leads to nonexistence of a BTZ-type black hole irrespective 
of symmetry breaking scale. The only spatial structure formed by the topological
lump is regular hyperboloid with deficit angle.

On the other hand, the asymptotic behavior of the nontopological solitons 
shows oscillation
around its boundary value $\pi/2$, and these solutions are
characterized by the number of nodes for a given parameter set of the model.
 The energy expressions of these
nontopological solitons include a logarithmic term at asymptotic region, and
this property resembles that of global U(1) vortices. According to the scale
of the negative cosmological constant, we obtained the following spacetimes:
One of them is regular hyperboloid with deficit angle and the other 
is charged BTZ black
hole. 
The conserved quasilocal
mass of the BTZ black hole is composed of two terms, i.e., one of them is
finite core mass and the other is logarithmically divergent term. 

Here we have several comments on some resemblance and difference between
our half integral winding $\sigma$ solitons and the global U(1) vortices. 
First, the
former solutions are nontopological, but the latter solutions are
topological. Therefore, the energetics of our nontopological solitons
should be checked to confirm their stability, which may provide a clue to
distinguish one from the other. 
Second, the global U(1) vortex is unique
regular soliton configuration with monotonically increasing scalar amplitude
for a given set of model parameters. On the other hand, a number of
nontopological solitons exist in a given model, which are characterized by the
number of oscillations in scalar amplitude.
Third, both solitons carry a long range term $(\sim 1/r^{2})$ in the
expressions of their energy density due to nontrivial phase winding sector of 
Goldstone modes. The solutions have been seen to tend towards black holes as
the symmetry breaking scale increases and the magnitude of negative
cosmological constant becomes small. The black hole generated by a
nontopological $\sigma$ soliton is a charged BTZ black hole without
non-Abelian scalar hair, 
while a small BTZ black hole lying within a global U(1)
vortex is available where nontrivial scalar field exists outside the horizon.

Since the
Einstein gravity in 2+1 dimensions does not have propagating degrees of
freedom, the introduction of a negative vacuum energy plays a drastic role for
making the soliton excitations rich in scalar theories. It made the global 
U(1) vortices
free from the physical curvature singularity in the model of a spontaneously
broken global U(1) symmetry. In our O(3) nonlinear $\sigma$ model this
attractive force supports the nontopological solitons, which have never
been obtained without adding
a gauge field and explicit symmetry breaking scalar potential \cite{GG}
except for some unstable, spherically symmetric solitons in (3+1)D de Sitter
spacetime \cite{AL}. 
The obtained spacetimes include charged
BTZ black hole. In this context it may also be intriguing to ask the same
question to local vortices in Abelian Higgs model \cite{Gar,Got}.
When we consider the stability of the obtained solutions or general straight
infinite cosmic strings, various forms of metric can also be taken into account,
e.g., a metric with boost invariance along the string direction,
$ds^2=e^{2N(r)}B(r)(dt^2-dz^2)-\frac{dr^2}{B(z)}-r^2d\theta^2$, or the general
form of static metric,
$ds^2=e^{2N(r)}B(r)(dt-C(r)dz)^2-\frac{dr^2}{B(z)}-r^2d\theta^2-D(r)dz^2$, or
even a stationary one,
$ds^2=e^{2N(r)}B(r)(dt-E(r)rd\theta)^2-\frac{dr^2}{B(z)}-r^2d\theta^2$.

Throughout this paper we have considered the cases where the deficit angle is
smaller than $2\pi$. If we recall that supermassive local vortices produced
various geometrical structures including an analog of Kasner spacetime, 
a cylinder, or a 
two sphere \cite{Got,LG}, we
may expect some drastic change of (anti-de Sitter) spacetime formed by the
topological lumps in the Planck scale. 
In relation with time-dependent soliton configurations, 
once stationary Q-lump solution is generated and  
forms a black hole structure \cite{Lee}, it must be a spinning
black hole in 2+1 dimensions.

\acknowledgments{
The authors would like to thank Chanju Kim and Kyoungtae Kimm for helpful 
discussions. This work was supported by the Ministry of Education 
(BSRI/97-2418),
the KOSEF (Grant No. 95-0702-04-01-3 and through CTP, SNU), 
and Faculty Research 
Fund, Sung Kyun Kwan University, 1997.}

\end{document}